\documentclass[sigconf,nonacm,balance=false,screen=true]{acmart}

\usepackage{subfigure}
\usepackage{array}
\usepackage{xurl}
\usepackage{algorithm}
\usepackage{algorithmic}

\setcopyright{none}
\AtBeginDocument{\hypersetup{colorlinks=true,linkcolor=blue,citecolor=blue,urlcolor=blue}}

\makeatletter
\renewcommand{\@mkauthors@iii}{%
  \gdef\@currentauthors{}%
  \gdef\@currentaffiliation{}%
  \global\setbox\mktitle@bx=\vbox{%
    \unvbox\mktitle@bx\par\medskip
    \centering
    {\fontsize{11}{15}\selectfont
      Hongyu Chen\textsuperscript{1}\quad
      Yikang Ruan\textsuperscript{1}\quad
      Yuhai Long\textsuperscript{2}\quad
      Ziyu Huang\textsuperscript{1}\quad
      Yiyu Luo\textsuperscript{1}\quad
      Liukun Yu\textsuperscript{1}\\
      Yangjie Zhou\textsuperscript{3,1}\quad
      Zihan Liu\textsuperscript{1}\quad
      Weihao Cui\textsuperscript{1}\quad
      Zijun Li\textsuperscript{1}\quad
      Anbang Wu\textsuperscript{1}\quad
      Yu Feng\textsuperscript{1}\\
      Jing Yang\textsuperscript{4,1}\quad
      Jingwen Leng\textsuperscript{1}\quad
      Bingsheng He\textsuperscript{3}\quad
      Minyi Guo\textsuperscript{1}\quad
      Shixuan Sun\textsuperscript{1,*}\par}
    \smallskip
    {\normalsize
      \textsuperscript{1}Shanghai Jiao Tong University\quad
      \textsuperscript{2}Sun Yat-sen University\\
      \textsuperscript{3}National University of Singapore\quad
      \textsuperscript{4}Guizhou University\par}
    \smallskip
    {\fontsize{8}{10}\selectfont
      \texttt{\{chenhongyu2048,yikang.ruan,huang\_ziyu,yiyu-luo,altair.liu\}@sjtu.edu.cn}\\
      \texttt{\{weihao,lzjzx1122,y-feng,leng-jw,sunshixuan\}@sjtu.edu.cn}\\
      \texttt{\{anbang,guo-my\}@cs.sjtu.edu.cn}\quad
      \texttt{\{yj\_zhou,dcsheb\}@nus.edu.sg}\\
      \texttt{longyh27@mail2.sysu.edu.cn}\quad
      \texttt{yu\_liukun@outlook.com}\quad
      \texttt{jyang23@gzu.edu.cn}\par}
    \medskip
  }%
}
\makeatother

\begin{document}

\title[LiquidCP: Decoupling Attention Tasks from Communication and Kernel Boundaries in Context Parallelism]{LiquidCP: Decoupling Attention Tasks from Communication and Kernel Boundaries in Context Parallelism}

\author{Hongyu Chen}
\email{chenhongyu2048@sjtu.edu.cn}
\affiliation{
  \institution{Shanghai Jiao Tong University}
  \city{Shanghai}
  \country{China}
}

\author{Yikang Ruan}
\email{yikang.ruan@sjtu.edu.cn}
\affiliation{
  \institution{Shanghai Jiao Tong University}
  \city{Shanghai}
  \country{China}
}

\author{Yuhai Long}
\email{longyh27@mail2.sysu.edu.cn}
\affiliation{
  \institution{Sun Yat-sen University}
  \city{Guangzhou}
  \country{China}
}

\author{Ziyu Huang}
\email{huang\_ziyu@sjtu.edu.cn}
\affiliation{
  \institution{Shanghai Jiao Tong University}
  \city{Shanghai}
  \country{China}
}

\author{Yiyu Luo}
\email{yiyu-luo@sjtu.edu.cn}
\affiliation{
  \institution{Shanghai Jiao Tong University}
  \city{Shanghai}
  \country{China}
}

\author{Liukun Yu}
\email{yu\_liukun@outlook.com}
\affiliation{
  \institution{Shanghai Jiao Tong University}
  \city{Shanghai}
  \country{China}
}

\author{Yangjie Zhou}
\email{yj\_zhou@nus.edu.sg}
\affiliation{
  \institution{National University of Singapore}
  \city{Singapore}
  \country{Singapore}
}
\affiliation{
  \institution{Shanghai Jiao Tong University}
  \city{Shanghai}
  \country{China}
}

\author{Zihan Liu}
\email{altair.liu@sjtu.edu.cn}
\affiliation{
  \institution{Shanghai Jiao Tong University}
  \city{Shanghai}
  \country{China}
}

\author{Weihao Cui}
\email{weihao@sjtu.edu.cn}
\affiliation{
  \institution{Shanghai Jiao Tong University}
  \city{Shanghai}
  \country{China}
}

\author{Zijun Li}
\email{lzjzx1122@sjtu.edu.cn}
\affiliation{
  \institution{Shanghai Jiao Tong University}
  \city{Shanghai}
  \country{China}
}

\author{Anbang Wu}
\email{anbang@cs.sjtu.edu.cn}
\affiliation{
  \institution{Shanghai Jiao Tong University}
  \city{Shanghai}
  \country{China}
}

\author{Yu Feng}
\email{y-feng@sjtu.edu.cn}
\affiliation{
  \institution{Shanghai Jiao Tong University}
  \city{Shanghai}
  \country{China}
}

\author{Jing Yang}
\email{jyang23@gzu.edu.cn}
\affiliation{
  \institution{Guizhou University}
  \city{Guiyang}
  \country{China}
}
\affiliation{
  \institution{Shanghai Jiao Tong University}
  \city{Shanghai}
  \country{China}
}

\author{Jingwen Leng}
\email{leng-jw@sjtu.edu.cn}
\affiliation{
  \institution{Shanghai Jiao Tong University}
  \city{Shanghai}
  \country{China}
}

\author{Bingsheng He}
\email{dcsheb@nus.edu.sg}
\affiliation{
  \institution{National University of Singapore}
  \city{Singapore}
  \country{Singapore}
}

\author{Minyi Guo}
\email{guo-my@cs.sjtu.edu.cn}
\affiliation{
  \institution{Shanghai Jiao Tong University}
  \city{Shanghai}
  \country{China}
}

\author{Shixuan Sun}
\authornote{Corresponding author.}
\email{sunshixuan@sjtu.edu.cn}
\affiliation{
  \institution{Shanghai Jiao Tong University}
  \city{Shanghai}
  \country{China}
}

\begin{abstract}
Context parallelism (CP) partitions long sequences across GPUs for LLM training and inference.
For variable-length training and full prefill, attention-balanced placement can leave token loads uneven, while ring execution repeats overhead at each computation and communication step. 
In stage-serialized decode CP (DCP), long-running tasks delay dependent work and leave substantial GPU hardware idle.
We propose LiquidCP, which unifies training and inference through a common model of Transfer, Compute, and Reduction tasks with tile-level dependencies, decoupling attention tasks from communication and kernel boundaries.
For training and full prefill, MegaRing incorporates BuddyRing placement to balance attention and token loads, and coalesces ready KV regions during forward attention to amortize overhead.
For decoding and mixed batches, MegaDCP selects history-KV splits by jointly modeling the completion of attention and its dependent reductions.
A shared persistent GPU-resident engine executes both designs.
Compared with the fastest baseline per configuration on eight H100 GPUs, MegaRing achieves geometric-mean core-attention speedups of $1.23\times$ forward and $1.18\times$ backward, and accelerates a Transformer layer by $1.08$--$1.16\times$.
MegaDCP achieves geometric-mean core-attention speedups of $1.20\times$ for decode-only and $1.35\times$ for mixed batches.
In vLLM serving with KV prefix cache, it achieves an average time-to-first-token speedup of $1.22\times$.
\end{abstract}

\keywords{Context Parallelism, Attention, Long-Context LLM, Machine Learning Systems}

\renewcommand{\shortauthors}{Hongyu Chen et al.}
\maketitle

\section{Introduction}
\label{sec:introduction}

Long-context training and inference impose substantial computation and memory demands on LLM systems.
Context parallelism (CP) addresses these demands by partitioning sequences across GPUs along the token dimension, distributing attention computation and the storage of activations and KV caches~\cite{ring_attention,megatron_hybrid_cp}.
In training and full prefill, query and KV sequences are both long.
Grouped-query attention (GQA)~\cite{gqa} uses fewer KV heads than query heads, favoring dataflows that keep queries local and exchange KV, typically around a ring~\cite{ring_attention,ring_flash_attention}.
In decoding and chunked prefill over cached contexts, queries can be much shorter than their KV histories.
Decode context parallelism (DCP) instead shards history KV and exchanges queries and partial attention states~\cite{vllm_dcp,helix_parallelism}.

Efficient CP requires an \emph{end-to-end view of load balance and computation--communication coordination}.
Placement determines attention work and token loads across GPUs, while task partitioning determines attention parallelism, per-task overhead, and reduction work.
Communication and reduction dependencies further determine when tasks become executable and when their outputs become usable.
Consequently, balancing aggregate workload alone does not ensure efficient execution: task boundaries and the progression of dependent work also matter.

\emph{Training and full prefill.}
Dynamic CP adapts placement and parallelism degrees to variable-length batches~\cite{megatron_hybrid_cp,zeppelin}.
FCP further combines computation- and memory-aware block placement with communication planning and stage coalescing~\cite{zhao2026unleashing}.
Attention and tokenwise operations have different scaling characteristics, making complete-layer performance important when assessing placement:
attention work grows approximately quadratically with sequence length, whereas many other operations scale with token count.
On the GitHub dataset, Megatron Hybrid CP reduces core-attention latency by 46.6\% relative to a ring CP baseline using FlashAttention-3 (FA3)~\cite{flashattention3}, while increasing the latency of the remaining Transformer-layer operations by 38.3\%.
Execution introduces another source of overhead.
The conventional ring baselines evaluated here invoke attention separately for each KV communication block, repeating setup and partial-result merging across steps (Section~\ref{sec:motivation}).

\emph{Decoding.}
Evenly sharded history KV can still yield uneven task completion within each GPU: short queries in decode-only batches limit query parallelism, while heterogeneous histories produce tasks of unequal duration. 
In mixed batches that combine decoding and chunked prefill requests, long prefill queries can cause FA3's batch-wide heuristic to leave decoding histories unsplit.
The FA3-based DCP paths of state-of-the-art inference frameworks~\cite{vllm_dcp,sglang_software} execute communication, attention, and reduction as separate dependent kernel stages, so a few long tasks can delay subsequent work.
Additional KV splits expose parallelism but also add task overhead and reduction work.
An effective policy must therefore consider attention together with its dependent reductions, while allowing completed portions of the workload to advance (Section~\ref{sec:motivation}).

\begin{figure}[t]
    \centering
    \includegraphics[width=0.85\columnwidth]{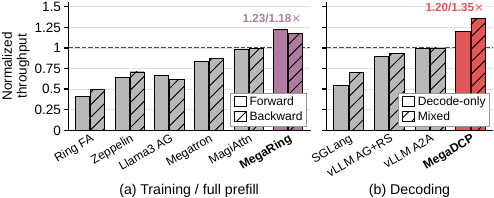}
    \vspace{-4pt}
    \caption{
        \textbf{Core-attention throughput on eight H100 GPUs.}
        Throughput is normalized to the best evaluated baseline in each configuration (dashed line); labels show LiquidCP's speedups.
    }
    \label{fig:intro-teaser}
    \vspace{-8pt}
\end{figure}

In this paper, we present \emph{LiquidCP}, which coordinates attention-task formation and partitioning with fine-grained fused execution for CP.
Building on composable partial attention states~\cite{ring_attention,ye2025flashinfer}, we describe execution using \emph{Transfer}, \emph{Compute}, and \emph{Reduction} tasks linked by tile-level dependencies (Section~\ref{sec:task_model}).
This formulation exposes the trade-off between coalescing tasks to amortize overhead and splitting tasks to expose parallelism.
LiquidCP develops policies for both regimes, accounting for input readiness and downstream reduction costs.

For training and full prefill, \emph{MegaRing} combines structured placement with readiness-driven task formation (Section~\ref{sec:megaring}).
Its placement algorithm, \emph{BuddyRing}, jointly selects sequence CP degrees and locations within aligned, nested GPU groups.
It searches for placements satisfying attention-work and token-load tolerances while prioritizing reduced fragmentation of short sequences.
MegaRing executes groups of different sizes together in one launch.
During forward attention, it retains fine-grained KV transfers while dynamically forming Compute tasks from ready, unassigned KV regions across communication blocks and data owners.
Computation granularity thus adapts to runtime input availability within prescribed bounds, enabling early computation while amortizing per-task overhead.

For decoding and mixed batches, \emph{MegaDCP} retains the existing history-KV placement and selects splits using a \emph{critical-wave model} (Section~\ref{sec:megadcp}).
The model predicts the completion of history attention together with its dependent reductions, accounting for the additional work introduced by splitting and the overlap permitted by the executor.

To realize these policies, we develop a shared GPU-resident engine that fuses communication, attention, and reductions within persistent kernels (Section~\ref{sec:engine}).
Building on prior persistent and fused execution techniques~\cite{hazy_megakernel,mpk,parallelkittens,mkernel}, the engine coordinates KV arrival, partial-state accumulation, and output completion through device-side protocols.
These mechanisms support readiness-driven forward task formation and allow dependent reductions and transfers to advance as their inputs become available, while unrelated attention tasks continue.

We evaluate LiquidCP on eight NVIDIA H100 GPUs against training CP implementations~\cite{ring_flash_attention,llama3,megatron_hybrid_cp,zeppelin,magiattention}
and the FA3-based DCP paths of vLLM and SGLang~\cite{vllm_dcp,sglang_software}, reporting speedups over the fastest evaluated baseline per configuration (Figure~\ref{fig:intro-teaser}).
Across five variable-length datasets, MegaRing achieves geometric-mean core-attention speedups of $1.23\times$ forward and $1.18\times$ backward, and accelerates a Qwen3-30B-A3B Transformer layer~\cite{qwen3} by $1.08$--$1.16\times$.
On Mooncake trace-derived workloads, MegaDCP achieves geometric-mean core-attention speedups of $1.20\times$ for decode-only and $1.35\times$ for mixed batches.
In vLLM serving with preloaded history KV, MegaDCP achieves average speedups of $1.04\times$ in time between tokens and $1.22\times$ in time to first token.
This paper makes three contributions:

\begin{itemize}
    \item
    \textbf{Placement and task formation for training.}
    BuddyRing searches structured CP-group placements under token- and attention-load tolerances, while MegaRing dynamically forms forward attention tasks from ready KV regions.

    \item
    \textbf{Reduction-aware scheduling for DCP.}
    MegaDCP selects history-KV splits by modeling the completion of attention and its dependent reductions together.

    \item
    \textbf{Fine-grained fused execution for CP.}
    A GPU-resident engine supports runtime task formation and coordinates attention, reductions, and transfers through CP-specific readiness and completion protocols.
\end{itemize}

\section{Background and Motivation}
\label{sec:background-motivation}

\subsection{Preliminaries}
\label{sec:background}

CP partitions attention along the token dimension, with dataflows determined by the relative query and KV lengths.
A \emph{partial attention state} contains a locally normalized output and the log-sum-exp (LSE) of scores over a KV subset.
States over disjoint subsets can be merged into the complete output.

\begin{figure}[t]
    \centering
    \subfigure[Ring-based CP]{%
        \includegraphics[width=0.48\columnwidth]
            {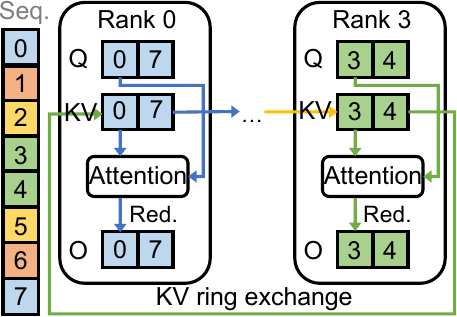}%
        \label{fig:ring-cp-dataflow}%
    }%
    \hfill
    \subfigure[DCP]{%
        \includegraphics[width=0.48\columnwidth]
            {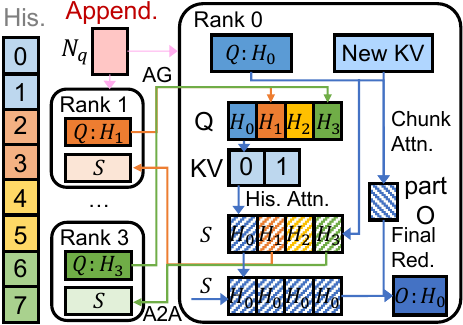}%
        \label{fig:dcp-dataflow}%
    }
    \vspace{-4pt}
    \caption{
        \textbf{CP dataflows across four ranks.}
        Numbers identify sequence or history-KV partitions;
        colors indicate owning ranks or query-head groups.
        (a) Ring CP with zigzag placement.
        (b) DCP with $\mathrm{TP}=\mathrm{DCP}=4$ and $H_{kv}=1$,
        viewed from Rank~0.
        $H_r$ denotes rank $r$'s query-head group;
        hatched $S$ blocks denote partial attention states.
        AG, A2A, and Red.\ denote all-gather, all-to-all,
        and LSE-weighted reduction.
    }
    \label{fig:cp-dataflows}
    \vspace{-8pt}
\end{figure}

\paragraph{Training and full prefill.}
For a sequence of length $N$, query and KV lengths satisfy
$N_q=N_{kv}=N$.
Grouped-query attention (GQA) uses fewer KV heads than query heads ($H_{kv}<H_q$), favoring KV exchange over exchanging queries and returning partial outputs.
Ring CP retains $N/P$ queries on each of $P$ ranks and circulates KV shards, computing and merging partial results at each step.
For causal attention, zigzag placement divides the sequence into $2P$ equal blocks and assigns blocks $p$ and $2P-1-p$ to rank $p$, balancing attention work (Figure~\ref{fig:ring-cp-dataflow}).
In backward, $dQ$ accumulates locally, while $dK$ and $dV$ contributions return to their KV owners.
For variable-length batches, dynamic CP~\cite{bytescale,megatron_hybrid_cp,zeppelin,magiattention} adjusts per-sequence parallel degrees and placement to avoid over-partitioning short sequences.

\paragraph{Decoding.}
Under tensor parallelism (TP) of degree $T$, each rank owns $h_q=H_q/T$ query heads.
When $H_{kv}<T$, head-wise partitioning replicates each KV head's cache across $T/H_{kv}$ ranks.
Decode context parallelism (DCP), supported by vLLM~\cite{vllm_dcp} and SGLang~\cite{sglang_software}, shards these replicated caches along the token dimension.
We consider groups of $P=T/H_{kv}$ ranks sharing one KV head.
For $N_q$ appended tokens and $N_{kv}$ history tokens, exchanging queries and partial states moves substantially less data than exchanging history KV when $N_qPh_q\ll N_{kv}$.

In the all-to-all dataflow shown in Figure~\ref{fig:dcp-dataflow}, each rank computes causal \emph{chunk attention} using its local queries and replicated appended KV.
Query all-gather enables \emph{history attention} for all group query heads against each rank's local history shard.
\emph{KV splitting} divides this shard into independent ranges, whose partial states are combined locally by \emph{history reduction}.
All-to-all returns these contributions to their query-head owners, where \emph{final reduction} merges them with the local chunk result.
Batches may contain only decoding requests (\emph{decode-only}) or also prefill chunks appended to cached contexts (\emph{mixed}).

\subsection{Motivation}
\label{sec:motivation}

\begin{figure}[tbp]
    \centering
    \includegraphics[width=\columnwidth]
        {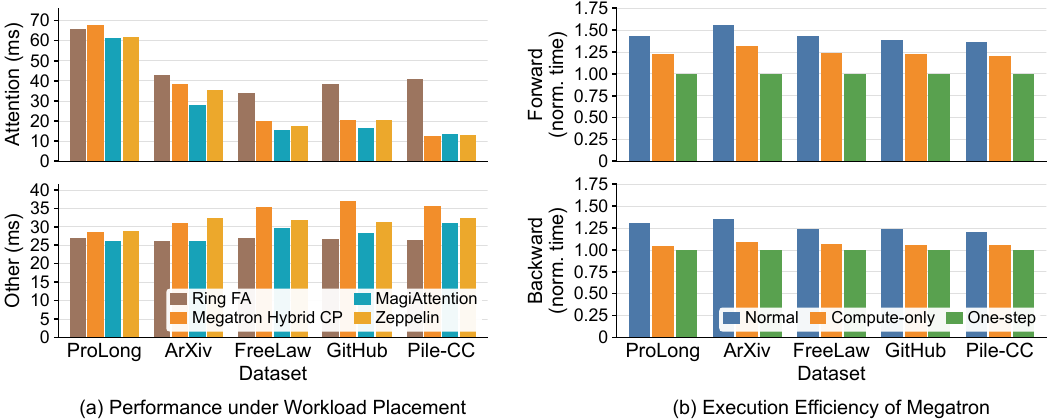}
    \vspace{-4pt}
    \caption{
        \textbf{CP placement and execution efficiency.}
        (a) Mean forward-plus-backward latency of core attention (top) and remaining layer operations (bottom).
        (b) Forward (top) and backward (bottom) latency under identical Megatron Hybrid CP assignments, normalized to One-step separately for each dataset and phase.
        Both panels use eight GPUs with a global 128K-token target.
    }
    \label{fig:cp-motivation}
    \vspace{-8pt}
\end{figure}

\paragraph{Attention and token load balance.}
Megatron Hybrid CP~\cite{megatron_hybrid_cp} initializes CP degrees from sequence lengths and places sequences by estimated attention work, without explicitly balancing per-rank token counts.
Figure~\ref{fig:cp-motivation}(a) separates layer latency into core attention (\emph{Attention}) and remaining operations (\emph{Other}).
On the GitHub dataset, Megatron Hybrid CP reduces Attention latency by 46.6\% relative to Ring FA, while increasing Other latency by 38.3\%.
Thus, attention gains can coexist with higher costs elsewhere in the layer.
Attention work grows approximately quadratically with sequence length, whereas tokenwise work grows linearly with token count; balancing one does not guarantee balancing the other.
This motivates a placement objective that accounts for both loads.

\paragraph{Communication and ring-step overhead.}
To separate execution effects from placement, we compare three modes under identical Megatron Hybrid CP assignments.
\emph{Normal} retains overlapped ring communication and per-step computation.
\emph{Compute-only} makes KV available before timing and excludes KV/dKV transfers from the timed region, preserving the same steps and local operations.
\emph{One-step} further combines the local attention work of sequences with the same CP degree into a single call.
In forward, removing transfers yields a 1.13--1.19$\times$ speedup, and merging steps yields a further 1.21--1.32$\times$ speedup (Figure~\ref{fig:cp-motivation}(b)).
The corresponding backward speedups are 1.15--1.25$\times$ and 1.05--1.10$\times$.
The first gap measures the aggregate impact of KV and gradients transfers under overlap.
The second persists with KV already available and combines changes in invocation count, partial-result merging, and computation shapes.

\textbf{\emph{Insight.}}
KV readiness constrains when computation can begin, but communication steps need not define its boundaries.
This motivates exposing inputs through fine-grained transfers while coalescing ready KV regions across communication steps to amortize task and merge overhead.

\begin{figure}[t]
    \centering
    \includegraphics[width=0.9\columnwidth]
        {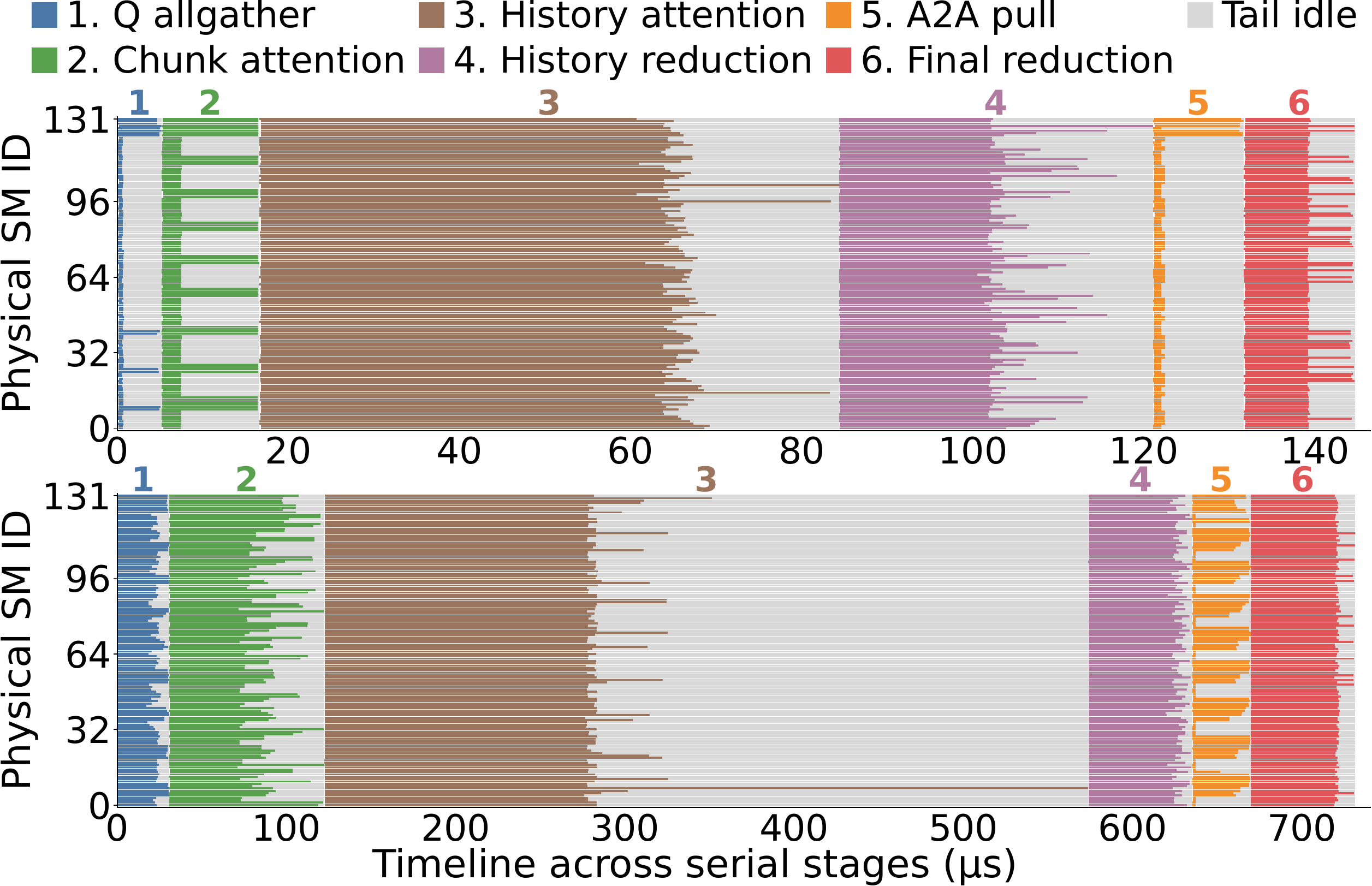}
    \vspace{-4pt}
    \caption{
        \textbf{Streaming multiprocessor (SM) activity across serialized DCP stages.}
        Timelines show a selected rank for decode-only (top) and mixed (bottom) batches on eight GPUs with
        $\mathrm{TP}=8$ and $\mathrm{DCP}=2$.
        Colors indicate stage work; gray indicates tail idle.
        The horizontal axis concatenates each stage's work span, from its earliest start to its latest completion.
    }
    \label{fig:dcp-sm-utilization-motivation}
    \vspace{-8pt}
\end{figure}

\paragraph{Execution tails in DCP.}
The vLLM DCP path executes communication, attention, and reduction as sequential kernels~\cite{vllm_dcp_ops}.
Consequently, early-finishing SMs cannot advance to later stages until the current stage's last task finishes.
We examine this pattern using a stage-serialized variant of our executor with native FA3 and explicit inter-stage barriers.
We define \emph{tail idle} as an SM's wait after its last task in a stage until that stage finishes.
Its fraction is summed tail idle divided by total SM-time across the plotted stages: 40.1\% for the decode-only batch and 48.1\% for the mixed batch (Figure~\ref{fig:dcp-sm-utilization-motivation}).

The native FA3 heuristic~\cite{flashattention3,flash_decoding,flashattention3_heuristic} uses the batch's maximum query length to bound KV splitting.
In the plotted mixed batch, long prefill queries set this bound to one, leaving long-history decoding requests unsplit.
Batch-wide query parallelism thus conceals long-running decoding tasks that delay stage completion.
In the decode-only batch, splitting exposes more parallel work, but a few last wave tasks still keep other SMs waiting.

\textbf{\emph{Insight.}}
KV splitting reduces work per attention task but adds partial results to merge.
This motivates selecting splits by the completion of history attention and its dependent reductions, while allowing ready consumers to proceed once their own inputs are available.

\section{A Common Task Model for Context-Parallel Attention}
\label{sec:task_model}

\begin{figure}[t]
    \centering
    \includegraphics[width=\columnwidth]{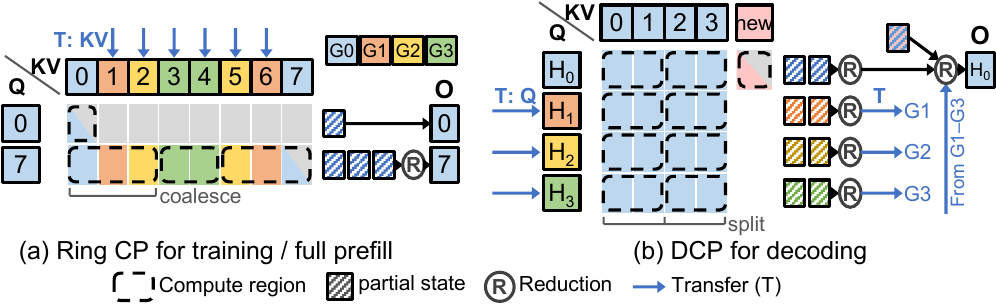}
    \vspace{-4pt}
    \caption{
        \textbf{Ring CP and DCP in the task model, viewed on GPU~0.}
        Rows denote Q tiles (query-head groups $H_r$ in DCP),
        columns denote KV tiles, and colors identify data owners.
        Gray cells are masked; ``new'' denotes appended KV.
        Arrows, dashed regions, and \textsf{R} nodes denote
        Transfer, Compute, and Reduction, respectively.
    }
    \label{fig:task-model}
    \vspace{-8pt}
\end{figure}

\paragraph{Attention Regions.}
An \emph{attention region} $A=I\times J$ consists of query tiles $I$ and KV tiles $J$, evaluated under the attention mask.
A region may contain noncontiguous tiles owned by different GPUs.
For data reuse, our implementation fixes one query tile per forward/decoding region and one KV tile per backward region.

For a query $q_i$, let $\{S_r\}$ partition its valid KV positions.
Let $o_i^{(r)}$ and $\operatorname{lse}_i^{(r)}$ denote the normalized attention output and LSE of scores over $S_r$.
Attention composition gives the complete output~\cite{flash_decoding}:
\begin{equation}
    o_i =
    \frac{
        \sum_r
        \exp\!\left(\operatorname{lse}_i^{(r)}\right)o_i^{(r)}
    }{
        \sum_r
        \exp\!\left(\operatorname{lse}_i^{(r)}\right)
    }.
    \label{eq:partitioned-attention}
\end{equation}
In backward, each region uses the complete forward outputs and LSEs, together with the output gradient $dO$, to produce additive contributions to $dQ$, $dK$, and $dV$.
If the regions cover every valid query–key interaction exactly once, summing their gradient contributions yields the complete attention gradients.

\paragraph{Primitive Operations.}
The model uses three logical primitives (Figure~\ref{fig:task-model}).
\textbf{Transfer} makes operand tiles or partial results available on their consuming GPU.
\textbf{Compute} evaluates one region and produces its partial results.
\textbf{Reduction} merges output/LSE states or sums gradient contributions for the same target, either incrementally or as a separate task.
These primitives may be fused within a single kernel.

Dependencies track the tiles and partial states consumed by each task.
A task may execute once its own inputs are available, independently of unrelated tasks.
A target becomes complete only after all required contributions have been combined.

\paragraph{Design Implications.}
Placement determines per-rank attention and token loads.
For a fixed placement, region size trades task and merge overhead against input readiness and task parallelism.
Coalescing amortizes costs, but waiting to form larger regions can delay dispatch and leave fewer tasks available for parallel execution.
Splitting exposes more tasks while creating more partial states and reduction work.

For long queries, MegaRing uses BuddyRing to balance attention and token loads through sequence placement, and coalesces ready KV regions across communication blocks in forward attention (Section~\ref{sec:megaring}).
For short queries over evenly sharded history KV, MegaDCP chooses splits by jointly modeling history attention and its dependent reductions (Section~\ref{sec:megadcp}).

\section{MegaRing for Training and Full Prefill}
\label{sec:megaring}

MegaRing transfers KV in fine-grained tiles and forms forward
Compute tasks by coalescing ready tiles across communication
blocks and KV owners.
Its placement algorithm, BuddyRing, balances token and attention
loads while limiting sequence fragmentation.

\subsection{Readiness-Driven Execution}
\label{sec:megaring-readiness}

\begin{figure}[tbp]
    \centering
    \includegraphics[width=0.95\linewidth]
        {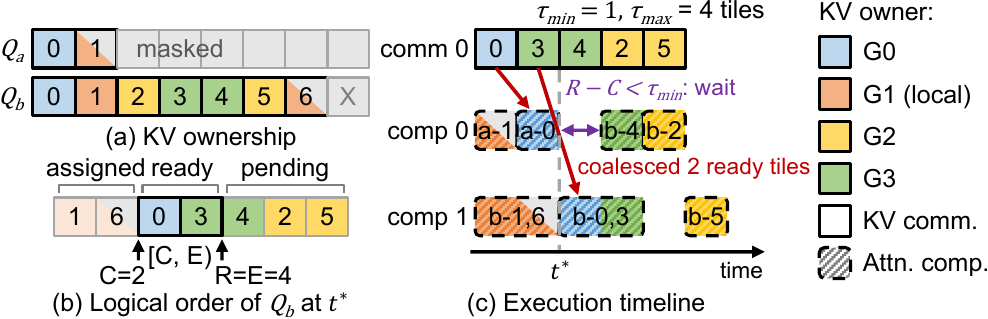}
    \vspace{-4pt}
    \caption{
        \textbf{Readiness-driven execution on GPU1 (G1)
        with four-GPU zigzag placement.}
        Colors identify KV owners; numbers denote KV tile indices.
        (a) Valid KV ranges for $Q_a$ and $Q_b$; gray regions are masked.
        (b) Logical KV order for $Q_b$ before task assignment at $t^*$.
        (c) Cooperative thread array (CTA) timelines with
        $\tau_{\min}=1$ and $\tau_{\max}=4$.
        Task labels identify the Q tile and processed KV tiles;
        arrows indicate dependencies.
    }
    \label{fig:megaring_forward_example}
    \vspace{-8pt}
\end{figure}

Given a placement, MegaRing indexes each Q tile's applicable
KV tiles in a logical order, with local tiles first and remote
tiles appended as their readiness is committed.
Let $R$ denote the endpoint of the ready prefix and $C$ the
endpoint of work already assigned to Compute tasks.
Thus, $[C,R)$ contains ready, unassigned KV tiles.
Communication CTAs extend the ready prefix, while compute CTAs
claim intervals from $[C,R)$.
After completing their transfers, communication CTAs join computation.

When $R-C\geq\tau_{\min}$, a compute CTA claims $[C,E)$, where
\begin{equation}
    E=C+\min(R-C,\tau_{\max}).
\end{equation}
Assignment advances $C$ to $E$ and fixes the task's interval;
later arrivals are handled by subsequent tasks.
Once all required KV tiles are ready, a final nonempty remainder
smaller than $\tau_{\min}$ may also be claimed.
Each task merges its partial output and LSE into the Q tile's
running state using Equation~\ref{eq:partitioned-attention}.
The output is complete only after all required KV contributions
have been merged.
Section~\ref{sec:engine-megaring} details readiness tracking,
task assignment, and result accumulation.

At $t^*$ in Figure~\ref{fig:megaring_forward_example},
$Q_b$ has $C=2$ and $R=4$.
Compute CTA~1 claims $[2,4)$, combining KV tiles~0 and~3
from G0 and G3 into one task.

Both thresholds are measured in KV tiles.
$\tau_{\min}$ trades earlier dispatch against amortization of
Q loading and partial-state updates;
$\tau_{\max}$ limits per-task work to preserve parallelism
when many tiles are ready.
For zigzag-partitioned sequences, we use the empirical defaults
\begin{equation}
    \tau_{\min}=\left\lceil\frac{H}{2}\right\rceil,
    \qquad
    \tau_{\max}=4H,
\end{equation}
where $H=L/(2gB_{\mathrm{KV}})$ is the number of KV tiles
in a zigzag half-shard.
Here, $L$ is the sequence length, $g$ its CP degree, and $B_{\mathrm{KV}}$ the number of tokens per KV tile. The thresholds are therefore sequence-specific.

Backward retains one KV tile per task, accumulating its
$dK$ and $dV$ in registers over the applicable local Q tiles,
which are already available.
Coalescing multiple KV tiles would require additional gradient accumulators and increase register pressure. We therefore leave backward coalescing to future work.

\subsection{Workload Placement with BuddyRing}
\label{sec:buddyring}

\paragraph{Placement space.}
Given sequence lengths $\{L_i\}$ and a power-of-two GPU count $N$,
BuddyRing assigns sequence $i$ to an aligned group $(s_i,g_i)$:
$g_i\in\{1,2,4,\ldots,N\}$ consecutive GPUs starting at $s_i$,
with $s_i\bmod g_i=0$.
These disjoint or nested groups form a buddy hierarchy.
Admissible group sizes and tile-alignment requirements are
specified in Appendix~\ref{app:buddyring}.
Sequences remain local when $g_i=1$ and use zigzag placement otherwise.
The groups of different sizes are executed in one MegaRing kernel launch.

\paragraph{Placement objective.}
Each group member receives $L_i/g_i$ tokens and estimated
attention work $a_i/g_i$, where $a_i=L_i(L_i+1)/2$
for causal attention.
Let $D_T$ and $D_W$ denote the maximum absolute relative
deviations of per-rank token and attention loads from their
respective means $\bar T$ and $\bar W$.
BuddyRing permits deviations up to $\epsilon_T$ and $\epsilon_W$
(10\% and 5\% by default), so small imbalances do not force
additional splitting.
A placement is feasible when its violation
\begin{equation}
    V=\max\left(0,\ D_T-\epsilon_T,\ D_W-\epsilon_W\right)
\end{equation}
is zero.

BuddyRing ranks placements lexicographically by $V$,
a splitting profile, the communication proxy
$\sum_i L_i(g_i-1)/g_i$, the number of distinct groups,
and finally $D_W$ and $D_T$.
The splitting profile compares, for each length class from
shortest to longest, the number of split sequences followed
by their summed split depths $\log_2 g_i$.
Once both load tolerances are satisfied, this ordering
prioritizes avoiding fragmentation of short sequences.

\paragraph{Search.}
BuddyRing uses large sequences to establish a placement and
smaller sequences to fill residual capacity.
Sequence $i$ is \emph{structural} if
$\kappa_i=\max(L_i/\bar T,a_i/\bar W)\geq0.5$;
the remaining sequences are \emph{fillers}.
A Pareto beam search places structural sequences in descending
$\kappa_i$ order, exploring their legal groups while retaining
partial placements with different load, fragmentation, and
communication trade-offs.
Fillers are initially placed greedily on single GPUs,
and the best completed placements undergo local refinement.

If no feasible placement is found, BuddyRing progressively
allows fillers to split, processing longer length classes first
and trying smaller groups before larger ones.
The search stops at the first feasible relaxation stage;
otherwise, it returns the best-ranked placement across all stages.
Appendix~\ref{app:buddyring} specifies candidate pruning,
filler placement, local refinement, and relaxation.
The planner runs once per batch and returns group assignments
ordered by decreasing group size to MegaRing's executor.

\begin{figure}[tbp]
    \centering
    \includegraphics[width=\columnwidth]
        {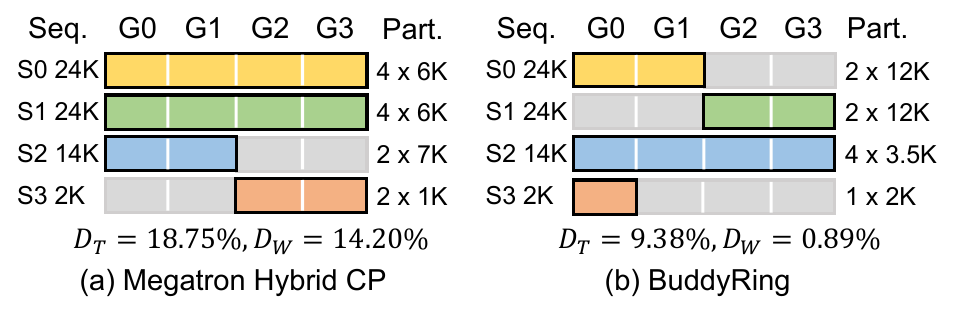}
    \vspace{-4pt}
    \caption{
        \textbf{Placement of a 64K-token batch on four GPUs.}
        Colors identify sequences; gray indicates GPUs outside
        each sequence's group.
        ``Part.'' denotes CP degree $\times$ tokens per GPU,
        with $\mathrm{K}=1024$.
        $D_T$ and $D_W$ report token and attention imbalance,
        respectively.
    }
    \label{fig:buddy_ring_example}
    \vspace{-8pt}
\end{figure}

In Figure~\ref{fig:buddy_ring_example}, BuddyRing keeps the
2K sequence local.
Megatron Hybrid CP splits it across two GPUs when expanding
groups initialized with an 8K-token per-GPU target.
BuddyRing's placement meets both load tolerances, reducing
$D_T$ from 18.75\% to 9.38\% and $D_W$ from 14.20\% to 0.89\%.
It uses nine sequence shards, compared with twelve for Megatron Hybrid CP, limiting fragmentation to preserve computational efficiency.

\section{MegaDCP for Decoding}
\label{sec:megadcp}

MegaDCP targets decode-only and mixed batches containing chunked prefill.
It keeps history KV evenly sharded and combines dependency-driven
execution with KV splitting guided by the joint completion of
attention and history reduction.

\subsection{Execution Model}
\label{sec:megadcp-execution}

On each GPU, MegaDCP executes the batch's DCP stages as a graph
of Transfer, Compute, and Reduction tasks within one persistent kernel.
The shared attention queue places local chunk tasks before history
tasks, allowing chunk attention to overlap query transfers.
A history-attention task becomes eligible when all its queries
are available; a history-reduction task becomes eligible when
all contributing partial attention results are ready.
Completed history contributions are published for transfer to
their query owners at output-block granularity.
Final reduction requires the block's local chunk and history
results and all remote history contributions.

Each compute CTA switches to reduction once all attention tasks are
claimed, allowing ready reductions to overlap with attention on other CTAs.
Section~\ref{sec:engine} describes task dispatch and
completion signaling.

\subsection{Critical-Wave KV Splitting}
\label{sec:megadcp_critical_wave}

\paragraph{Critical-wave model.}
Splitting changes both when partial results become ready and
when attention CTAs release warps for reduction.
MegaDCP's \emph{critical-wave model} jointly simulates attention
and history reduction on a GPU.

One time unit is the per-CTA time to process a history KV tile
with all $K$ compute CTAs executing attention.
With $k$ active attention CTAs, the modeled per-tile time is
\begin{equation}
    \tau(k)=\max\left(\rho,\,k/K\right).
\end{equation}
The $k/K$ term models shared HBM bandwidth, while $\rho$
imposes a calibrated per-tile time floor.
An attention task $t$ carries
$c_{\mathrm{attn}}(t)=n_t+\delta_{\mathrm{attn}}$
tile-equivalents of work, comprising $n_t$ KV tiles and
a fixed per-task work overhead $\delta_{\mathrm{attn}}$.
The simulator advances each running attention task at rate
$1/\tau(k)$, updating $k$ whenever an attention task starts
or finishes.

History-reduction duration is approximated by the
load-independent cost
\begin{equation}
    c_{\mathrm{red}}(u)
    =\delta_{\mathrm{red}}+\gamma v_u s_u.
\end{equation}
Here, $v_u$ counts output vectors (query token--head pairs),
$s_u$ counts contributing splits per vector,
$\delta_{\mathrm{red}}$ is the per-task overhead, and
$\gamma$ is the cost per partial vector, in normalized time units.

The simulator dispatches each candidate plan's attention tasks in order
to the first available compute CTA.
Here, $k$ counts CTAs executing chunk or history attention.
A CTA releases its warps to history reduction after completing
its last attention task.
Each reduction uses one warp and starts in queue order when its warp
and inputs are available.
The score is the latest simulated completion time across attention
and history-reduction tasks; lower is better.
The model omits query transfers, result exchange, and final reduction,
so its score serves as a proxy for kernel latency.

\paragraph{Task ordering.}
Decode-only plans with splitting use release-aware
longest-processing-time-first (LPT) ordering.
Query transfers are prioritized by the attention work they unlock.
History tasks are ordered by the transfer epoch of their last
required query block, then by decreasing $c_{\mathrm{attn}}$.
Unsplit and mixed-batch plans retain the original FIFO order of
attention tasks in the shared chunk/history queue.

\paragraph{Plan selection.}
The planner searches over per-sequence split counts, starting
from \emph{NoSplit}: one Compute task per query tile over
its full local history.
MegaDCP uses the native FA3 heuristic's split counts as the
\emph{FA3-derived} reference plan and initial per-sequence search
upper bounds.
For decode requests with long histories, MegaDCP raises these bounds
to at least four, preserving splitting options in mixed batches.

At each iteration, the planner evaluates increasing the split
count of each sequence below its limit by one.
It also evaluates joint increments for sequences whose tasks
tie at the attention tail, since changing one sequence alone
may leave the score unchanged.
The best-scoring candidate is accepted only if it improves
the current score.
Search stops when no candidate improves the score or no
further splitting is allowed.

MegaDCP selects the best-scoring plan among NoSplit, the search
result, and the FA3-derived reference.
A split plan is adopted only if its modeled completion time
is at least 10\% lower than NoSplit.
Appendix~\ref{app:critical-wave-params} specifies parameter
calibration, split limits, tie-breaking, and the construction
of transfer epochs.

\begin{figure}[tbp]
    \centering
    \includegraphics[width=0.9\columnwidth]
        {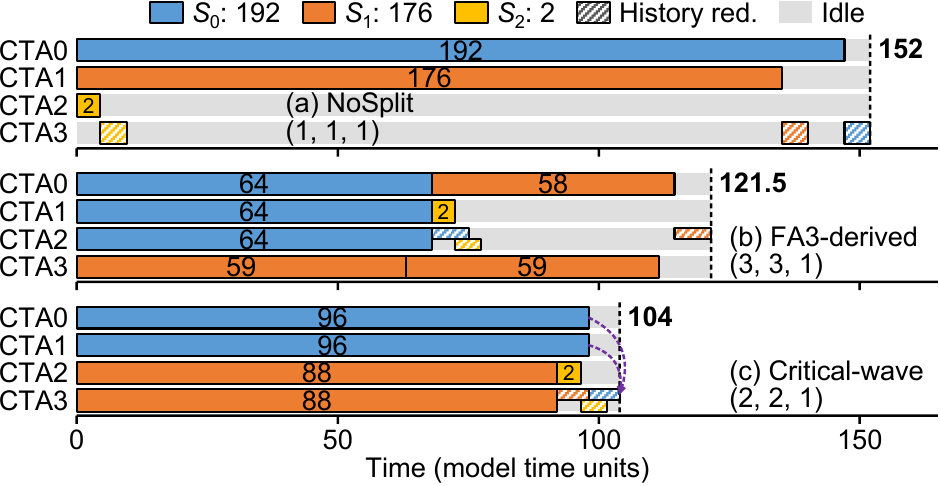}
    \vspace{-6pt}
    \caption{
        \textbf{Illustrative split plans under the critical-wave model.}
        Tuples list per-sequence KV partition counts; $(1,1,1)$ denotes NoSplit.
        Numbers inside attention blocks label KV-tile counts; hatched blocks are history
        reductions, drawn on the CTAs whose warps execute them.
        Block lengths reflect concurrency-dependent processing rates.
        Gray marks idle time, dashed lines mark modeled completion,
        and arrows show reduction dependencies.
        Illustrative parameters:
        $\delta_{\mathrm{attn}}=\delta_{\mathrm{red}}=4$,
        $\gamma=1$, $\rho=3/4$, and $v_u=1$.
    }
    \label{fig:critical_wave_example}
    \vspace{-16pt}
\end{figure}

Figure~\ref{fig:critical_wave_example} compares plans for
local histories of 192, 176, and 2 KV tiles on four compute CTAs.
Tuples give their partition counts in this order.
NoSplit $(1,1,1)$ leaves the long histories on two CTAs and finishes
at 152 model time units.
The FA3-derived plan $(3,3,1)$ exposes more parallelism but
creates a second wave of long tasks and more reduction work,
finishing at 121.5.
The selected plan $(2,2,1)$ distributes the long histories
across four concurrent tasks while limiting task and reduction
overhead, finishing at 104.

\section{Execution Engine Implementation}
\label{sec:engine}

MegaRing and MegaDCP share a GPU-resident engine that implements
the task primitives of Section~\ref{sec:task_model} using the
persistent execution structure of prior megakernels~\cite{hazy_megakernel,mpk}.
The host supplies placement and task metadata; CTAs initially
specialize in communication or computation and coordinate through
readiness flags and completion counters.
Appendix~\ref{app:engine-details} details the publication protocols
and CTA configurations.

\subsection{MegaRing Specialization}
\label{sec:engine-megaring}

Communication CTAs pull remote KV tiles directly from their owners
over NVLink into a local buffer, so each tile is transferred once and
reused by all local Q tiles and query heads.
MegaRing exposes out-of-order KV arrivals through a per-sequence
committed-prefix counter and a logical-to-physical KV map.
Only fully written tiles enter this contiguous, monotonic prefix.
For causal zigzag placement, a barrier ensures that all KV needed
by front-half queries is committed before publishing KV needed
only by back-half queries.
Compute CTAs combine the applicable remote prefix with local KV
to claim eligible intervals
(Section~\ref{sec:megaring-readiness}).
Forward epilogues merge partial outputs and LSE under a
per-Q-tile lock; a separate merged-KV counter tracks completion.
Communication CTAs join the compute scheduler after completing
their transfers.

Backward accumulates $dQ$ locally and stages $dK$ and $dV$ for their KV owners.
After local completion, communication CTAs apply Tensor Memory Accelerator (TMA) reduce-add operations to the owners' gradient accumulators and publish system-scope completion signals once these operations finish.
Owners wait for all expected signals before final gradient conversion.

\subsection{MegaDCP Specialization}
\label{sec:engine-megadcp}

MegaDCP dispatches host-prepared descriptors from a unified
chunk/history attention queue and separate history- and
final-reduction queues.
After its last attention task, each compute CTA synchronizes,
reuses its shared memory, and repartitions registers for reduction.

Communication CTAs pull query blocks and update local readiness
counters.
History attention waits for its required query blocks, and
attention tasks publish completion flags after writing partial
outputs and LSE to local GPU memory.
History reductions wait for their inputs and track completion
per destination and output block.
The last contributor to a remote output block notifies the query
owner, which pulls the results and records receive completion
after the transfer finishes.
Final reduction waits for local chunk-attention flags,
the local history-reduction counter, and all remote
receive-completion flags to indicate that its enclosing output
block's inputs are complete.

\section{Evaluation}
\label{sec:evaluation}

\subsection{Experimental Setup}
\label{sec:eval_setup}

\paragraph{Platform and workloads.}
All experiments run on one node with eight NVIDIA H100 SXM GPUs,
each with 80\,GB HBM and 450\,GB/s NVLink bandwidth per direction.
MegaRing uses causal attention with 32 query heads, eight KV heads,
and head dimension 128~\cite{llama3,qwen3}.
We sample 20 batches per dataset and token budget
(64K, 128K, and 256K) from ProLong~\cite{prolong},
ArXiv, FreeLaw, GitHub, and Pile-CC~\cite{the_pile},
covering distributions dominated by long or short sequences
(Figure~\ref{fig:workload-distributions}(a)).

For MegaDCP, offline scheduling with prefix caching generates
300 snapshots from the Mooncake trace~\cite{mooncake,mooncake_trace}
at arrival-rate scales of $1\times$, $2\times$, and $4\times$.
Replaying each snapshot at DCP degrees 2, 4, and 8 yields
630 decode-only and 270 mixed cases.
Each decoding request contributes 16 query tokens per iteration, one target and 15 draft tokens as in speculative decoding~\cite{speculative_decoding,deepseek_v3};
chunked prefill processes uncached suffixes in chunks of up to
4K tokens.
Appendix~\ref{app:eval_details} provides additional setup details
and workload statistics.

\paragraph{Baselines.}
MegaRing baselines include zigzag NCCL Ring
Attention~\cite{ring_attention,ring_flash_attention,nccl};
overlapped KV all-gather per sequence or across packed batches
(Llama-3 style~\cite{llama3});
the placement algorithms of Megatron Hybrid CP~\cite{megatron_hybrid_cp} and Zeppelin~\cite{zeppelin}, reimplemented with FA3/NCCL;
and MagiAttention's public variable-length API~\cite{magiattention}.
\emph{MegaRing All CP} zigzag-partitions every sequence across
all GPUs; \emph{MegaRing Hybrid} uses BuddyRing placement.

MegaDCP baselines reproduce the DCP paths of vLLM and SGLang on a shared FA3~\cite{flashattention3} attention kernel with Q all-gather.
vLLM-AG-RS~\cite{vllm,vllm_dcp_ops},
vLLM-A2A~\cite{vllm_dcp,helix_parallelism}, and
SGLang-AG-AR~\cite{sglang,sglang_software}
use output reduce-scatter, packed output/LSE all-to-all,
and output all-reduce, respectively.
SGLang communicates in FP32, whereas vLLM and MegaDCP use BF16.
We evaluate eager execution and CUDA Graph replay,
reporting Graph results unless noted otherwise.

\paragraph{Measurement.}
Core-attention timings include CP communication and result reduction
and use the slowest rank's execution time.
Throughput is effective attention FLOPs per second per GPU;
workload-weighted throughput divides total FLOPs by total time
and GPU count.
Case latency is the mean over 40 timed iterations for MegaRing
and the median over 60 for MegaDCP.
Appendix~\ref{app:eval_details} details the warm-up and
measurement procedures.
For variable-length workloads, we select the best
communication-CTA count from $\{4,8,12,16\}$ per configuration.
Core-attention timings exclude host-side planning, which runs once per batch on the CPU and is reused by every layer. 
MegaDCP’s planner, including descriptor packing, takes under 3~ms at p95 per decoding step and is included in the end-to-end serving results; BuddyRing depends only on sequence lengths, so it can run in the data loader ahead of each training step.

\subsection{Core Attention Performance}
\label{sec:eval_core}

\begin{figure}[t]
  \centering
  \includegraphics[width=\columnwidth]{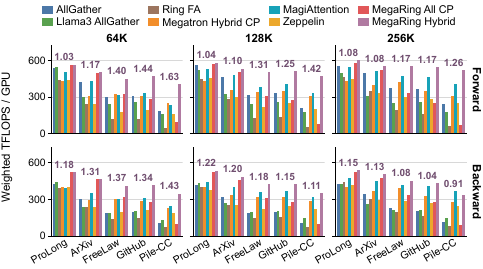}
  \vspace{-6pt}
  \caption{Workload-weighted forward and backward attention
  throughput per GPU. Hybrid annotations show speedup over the
  best baseline per dataset and token budget.}
  \label{fig:eval_megaring_variable}
    \vspace{-10pt}
\end{figure}

\paragraph{MegaRing.}
\label{sec:eval_megaring_variable}
MegaRing Hybrid outperforms the best baseline in all 15 forward
configurations and 14 of 15 backward configurations, with
geometric-mean speedups of 1.23$\times$ and 1.18$\times$,
respectively (Figure~\ref{fig:eval_megaring_variable}).
On ProLong and ArXiv, All CP already matches or exceeds the
best baseline, and BuddyRing adds at most 7\%.
Placement matters more on FreeLaw, GitHub, and Pile-CC,
where splitting short sequences across all eight GPUs leaves
too little work per task.
All CP falls to 0.17$\times$ the best baseline on Pile-CC,
whereas Hybrid reaches 1.63$\times$ in forward at 64K.
Gains diminish at larger token budgets as baselines better
amortize per-step costs.
Backward benefits less on these short-sequence-heavy datasets
because fixed KV tiles prevent coalescing of small ready regions.
On Pile-CC backward at 256K, Hybrid's latency is
8.8\% higher than MagiAttention's.

\begin{figure}[t]
  \centering
  \includegraphics[width=0.95\columnwidth]{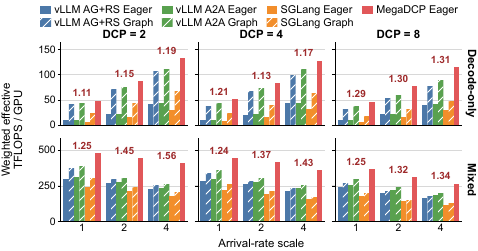}
  \vspace{-6pt}
  \caption{Workload-weighted attention throughput per GPU on decode-only (top) and mixed (bottom) batches.
  MegaDCP uses eager execution; baselines are evaluated with both eager execution and CUDA Graph replay.
  Annotations compare MegaDCP with the fastest CUDA Graph baseline.}
  \label{fig:eval_megadcp_variable}
    \vspace{-10pt}
\end{figure}

\paragraph{MegaDCP.}
\label{sec:eval_megadcp_variable}
MegaDCP achieves geometric-mean speedups of 1.20$\times$ on
decode-only batches and 1.35$\times$ on mixed batches over the
best CUDA Graph baseline
(Figure~\ref{fig:eval_megadcp_variable}).
Decode-only speedups increase from 1.11--1.19$\times$ at
DCP degree 2 to 1.29--1.31$\times$ at degree 8:
shorter local histories increase the relative cost of transfers
and reductions, making inter-task overlap more valuable.
Mixed-batch speedups grow with arrival rate, reaching
1.56$\times$ at degree 2 and a $4\times$ rate.
Section~\ref{sec:eval_megadcp_ablation_timestamp} examines
the contributions of planning and cross-stage execution.

Uniform-length MegaRing and uniform-history MegaDCP results
appear in Appendices~\ref{app:eval_uniform_megaring}
and~\ref{app:eval_uniform_megadcp}.

\subsection{Sources of MegaRing Gains}
\label{sec:eval_megaring_ablation}

\begin{figure}[t]
  \centering
  \subfigure[Execution ablation]{%
    \includegraphics[width=0.485\columnwidth]{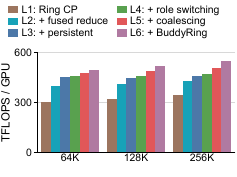}%
    \label{fig:eval_megaring_ablation}%
  }%
  \hfill
  \subfigure[Placement comparison]{%
    \includegraphics[width=0.485\columnwidth]{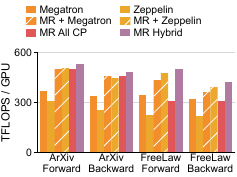}%
    \label{fig:eval_megaring_load_balance}%
  }
  \vspace{-6pt}
  \caption{MegaRing ablations.
  (a) Mean forward throughput on ArXiv at 64K, 128K, and 256K.
  (b) Workload-weighted forward/backward throughput on ArXiv
    and FreeLaw at 128K, comparing placements on the same MegaRing kernel with the baselines' FA3/NCCL execution.
  MR denotes MegaRing.}
  \label{fig:eval_megaring_sources}
    \vspace{-10pt}
\end{figure}

\paragraph{Execution mechanisms.}
Starting from ring CP with external reductions (L1),
L2--L5 cumulatively add fused reductions, persistent execution
with a fixed queue, CTA role switching, and readiness-driven
coalescing, while retaining All-CP placement
(Figure~\ref{fig:eval_megaring_ablation}).
Their incremental gains are 1.25--1.31$\times$,
6.1--12.9\%, 1.2--2.8\%, and 4.0--7.5\%, respectively.
Fusing reductions yields a substantial improvement;
coalescing further amortizes Q loads and partial-result updates
across larger tasks.
L6 adds BuddyRing placement, improving mean throughput by another 3--8\%.

\paragraph{Placement.}
\label{sec:eval_buddyring_placement}
We compare BuddyRing with Megatron-derived and Zeppelin-derived
placements using the same MegaRing kernel and sampled batches
at 128K tokens.
Across all five datasets, BuddyRing achieves geometric-mean
forward/backward speedups of 1.11/1.13$\times$ over
Megatron-derived placement and 1.05/1.07$\times$ over
Zeppelin-derived placement, with gains on every dataset.
Figure~\ref{fig:eval_megaring_load_balance} shows ArXiv and FreeLaw;
the larger gains on FreeLaw support balancing token and attention
loads while limiting short-sequence partitioning.
BuddyRing meets the 10\% token-load and 5\% attention-load tolerances in all 100 batches. On FreeLaw, mean per-batch $(D_T,D_W)$ values are $(6.4\%,4.3\%)$ for BuddyRing, $(44.8\%,33.3\%)$ for Megatron-derived, and $(18.7\%,6.8\%)$ for Zeppelin-derived placement.
Planner details are in Appendix~\ref{app:eval_details}.

\subsection{Sources of MegaDCP Gains}
\label{sec:eval_megadcp_ablation_timestamp}

\begin{figure}[t]
  \centering
  \includegraphics[width=0.95\columnwidth]{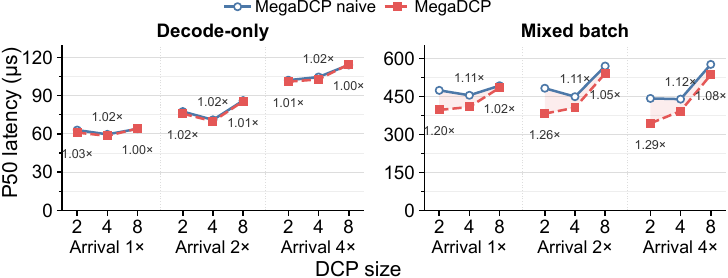}
    \vspace{-6pt}
    \caption{Mean per-case median kernel latency on decode-only (left) and mixed (right) batches. Annotations give the ratio of mean latencies, MegaDCP naive over MegaDCP.}
  \label{fig:eval_megadcp_critical_wave}
    \vspace{-10pt}
\end{figure}

\paragraph{Split planning and ordering.}
We compare MegaDCP with FA3-derived splits and FIFO ordering on the same engine (\emph{MegaDCP naive}).
Critical-wave planning and release-aware ordering provide 1.00--1.03$\times$ speedups on decode-only batches.
For mixed batches, both variants retain FIFO ordering, and critical-wave split planning provides 1.02--1.29$\times$ speedups (Figure~\ref{fig:eval_megadcp_critical_wave}).
For decode-only batches, both variants already split histories and overlap attention with reduction, leaving little additional benefit.
For mixed batches, the heuristic leaves long-history decoding
requests unsplit alongside prefill chunks.
Shortening these tasks gives the largest gain at DCP degree 2
and a $4\times$ arrival rate, where local histories are longest.

\paragraph{Cross-stage execution.}
At DCP degree 4 and a $4\times$ arrival rate,
MegaDCP reduces mean attention-pipeline latency from
128 to 107\,$\mu$s on decode-only batches and from
570 to 394\,$\mu$s on mixed batches relative to vLLM-A2A Graph.
Stage profiling shows earlier query-transfer and result-exchange
completion, consistent with processing query and output blocks
as their dependencies become ready.
Appendix~\ref{app:eval_megadcp_timestamps} provides the timelines
and profiling methodology.

\subsection{End-to-End Performance}
\label{sec:eval_end_to_end}

\begin{figure}[t]
  \centering
  \includegraphics[width=0.95\columnwidth]{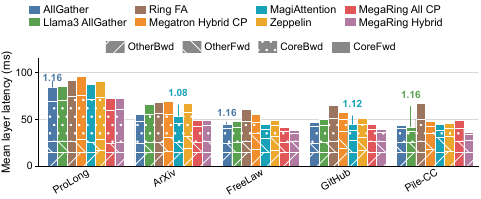}
    \vspace{-6pt}
    \caption{Transformer-layer latency averaged over 20 cases per dataset, decomposed into forward/backward and Core Attention/Other operations.
    Each MegaRing variant uses its best communication-CTA count per dataset. Annotations above the fastest non-MegaRing baseline show Hybrid's speedup.
    Zeppelin denotes our placement-derived baseline without linear-module remapping.}
  \label{fig:eval_megaring_end2end}
    \vspace{-10pt}
\end{figure}

\begin{figure}[t]
  \centering
  \includegraphics[width=0.95\columnwidth]{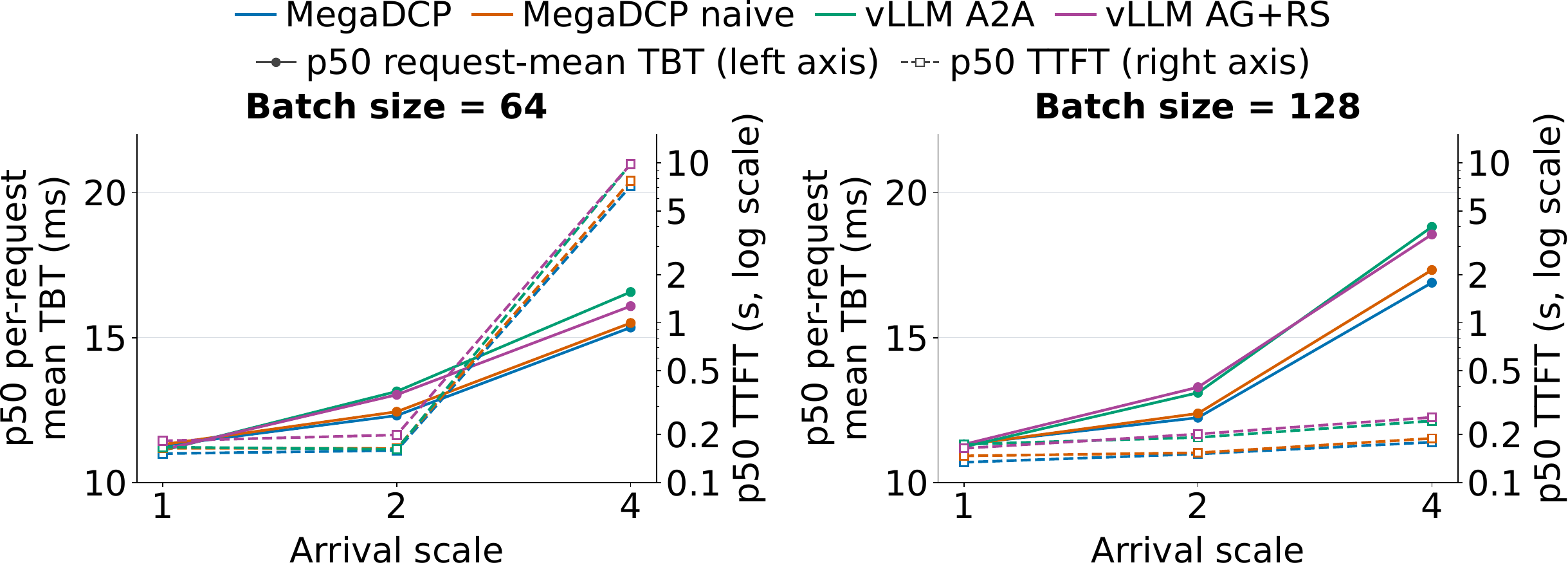}
  \vspace{-6pt}
  \caption{vLLM serving performance across arrival-rate scales
  and maximum batch sizes.
  Solid curves show median per-request mean TBT;
  dashed curves show median TTFT.
  All backends use CUDA Graphs.}
  \label{fig:eval_megadcp_end2end}
    \vspace{-10pt}
\end{figure}

We integrate both designs into workloads using
Qwen3-30B-A3B~\cite{qwen3};
Appendix~\ref{app:eval_end_to_end} gives the setup and further analysis.
For Transformer-layer training at a 128K token budget,
MegaRing Hybrid achieves 1.08--1.16$\times$ forward-plus-backward
speedups across the five datasets
(Figure~\ref{fig:eval_megaring_end2end}).
Faster attention and balanced token loads preserve the gains
across the remaining layer operations.

We evaluate vLLM serving with preloaded history KV and single-token decoding on a Mooncake trace segment.
Relative to the best vLLM baseline per configuration and metric, MegaDCP achieves 0.99--1.10$\times$ speedups in median per-request mean time between tokens (TBT) and 1.02--1.37$\times$ in median time to first token (TTFT) (Figure~\ref{fig:eval_megadcp_end2end}).
TBT gains are limited by unchanged non-attention computation; TTFT benefits more from the attention gains on mixed batches.

\section{Related Work}
\label{sec:relative_works}

\paragraph{Context parallelism and placement.}
\label{sec:related_cp}
CP uses ring KV exchange, KV all-gather, or head
redistribution~\cite{ring_attention,llama3,ulysses}, with hybrid
dataflows~\cite{usp,loongtrain} and improvements in overlap and causal
balance~\cite{distflashattn,striped_attention}.
Serving systems adapt parallelism or shard history
KV~\cite{loongserve,nanocp,vllm_dcp,sglang_software,helix_parallelism}.
Variable-length training uses adaptive partitioning, placement, and
block mapping~\cite{bytescale,hydraulis,megatron_hybrid_cp,training_dcp};
Zeppelin~\cite{zeppelin} remaps between attention and linear-module
layouts.
FCP~\cite{zhao2026unleashing} balances computation and memory through fixed-size block
placement and flexible peer-to-peer communication.
BuddyRing jointly selects sequence CP degrees and aligned power-of-two
groups, seeking a single layout that meets token- and attention-load
tolerances while limiting short-sequence fragmentation.

\paragraph{Attention task granularity.}
\label{sec:related_partitioning}
Single-GPU attention uses tiling, KV splitting, and composable partial
states~\cite{flashattention,flashattention3,flash_decoding,flashattention3_heuristic,ye2025flashinfer}.
Across GPUs, MagiAttention~\cite{magiattention} combines balanced
dispatch with multi-stage overlap, while UltraAttn~\cite{ultraattn}
schedules hierarchically tiled attention.
FCP~\cite{zhao2026unleashing} coalesces planned communication and computation
sub-stages, separating scheduling granularity from stage-execution
granularity.
MegaRing retains fine-grained KV transfers and dynamically forms
forward Compute tasks from ready, unassigned KV regions across
communication blocks and owners.
MegaDCP selects history-KV splits by jointly modeling attention and
dependent reductions, and orders tasks to advance downstream work.

\paragraph{GPU-resident execution.}
\label{sec:related_gpu_execution}
Megakernels provide persistent execution and dependency
management~\cite{hazy_megakernel,mpk,event_tensor}.
ParallelKittens~\cite{parallelkittens} and mKernel~\cite{mkernel}
fuse communication and computation, including Ring Attention;
UniEP~\cite{uniep} targets expert-parallel mixture-of-experts (MoE) training.
LiquidCP specializes these mechanisms for out-of-order KV arrival,
partial-state accumulation, and output-block completion, supporting
its task policies and fine-grained progression through attention,
reductions, and transfers.
\section{Conclusion}
\label{sec:conclusion}

This paper presented LiquidCP, which decouples attention tasks from communication and kernel boundaries in context parallelism.
Its common task model of Transfer, Compute, and Reduction primitives places training/full prefill and decoding at opposite ends of one design space: for long queries, MegaRing coalesces ready KV regions and BuddyRing balances attention and token loads, while for short queries, MegaDCP splits history KV by jointly modeling attention and its dependent reductions.
On eight H100 GPUs, MegaRing improves core-attention throughput on variable-length training batches by $1.23\times$ (forward) and $1.18\times$ (backward), and MegaDCP by $1.20\times$ on decode-only and $1.35\times$ on mixed batches, in geometric mean over the best baseline per configuration.

\bibliography{example_paper}
\bibliographystyle{ACM-Reference-Format}
\clearpage
\appendix
\setlength{\parskip}{2.5pt plus 1pt minus 0.5pt}
\setlength{\textfloatsep}{6pt plus 2pt minus 1pt}
\setlength{\intextsep}{6pt plus 1pt minus 1pt}
\setlength{\floatsep}{6pt plus 1pt minus 1pt}

\section{Technical Details}

\subsection{BuddyRing Search Details}
\label{app:buddyring}
 
\paragraph{Candidate groups.}
BuddyRing tests power-of-two group sizes up to $\min(N,8)$.
Causal splits require $256g_i$ to divide $L_i$, keeping all $2g_i$ zigzag chunks aligned to 128 tokens.
For a structural sequence, the minimum group size is the smallest legal $g$ satisfying $L_i/g\leq(1+\epsilon_T)\bar T$ and $a_i/g\leq(1+\epsilon_W)\bar W$, or the largest legal size if none qualifies.
The splitting profile in Section~\ref{sec:buddyring} uses length cutoffs of 2K, 4K, 8K, and 16K tokens.
 
\begin{table}[htb]
  \centering
  \vspace{-5pt}
  \caption{Default BuddyRing parameters.}
  \label{tab:buddyring_params}
  \small
  \setlength{\tabcolsep}{4pt}
  \begin{tabular*}{\columnwidth}{@{\extracolsep{\fill}}lr@{}}
    \toprule
    Parameter & Value \\
    \midrule
    Load tolerances $\epsilon_T$ / $\epsilon_W$ & 10\% / 5\% \\
    Structural threshold on $\kappa_i$ & 0.5 \\
    Beam width & 64 \\
    State-merging bin (of mean load) & 2\% \\
    Filler smooth-max $\lambda$ & 8 \\
    Top-ranked placements refined & 8 \\
    Maximum repair iterations & 32 \\
    GPUs per load extreme / max.\ sequences per GPU & 2 / 4 \\
    \bottomrule
  \end{tabular*}
\end{table}
 
\paragraph{Pareto beam search.}
For each structural sequence, the beam tries every allowed group for each retained placement and discards dominated candidates.
For load $X\in\{T,W\}$, overload is $\max(0,\max_r X_r/\bar X-1-\epsilon_X)$ and spread is $(\max_r X_r-\min_r X_r)/\bar X$.
For matching per-GPU load bins, BuddyRing keeps the best placement by splitting profile, communication proxy, then distinct group count.
The beam keeps the best candidate for each of these three criteria, each spread, and the larger overload.
It fills remaining slots by decreasing NSGA-II crowding distance over all criteria~\cite{deb2002nsga2}.
 
\paragraph{Filler placement.}
In decreasing $\kappa_i$ order, each filler goes to the GPU that minimizes the larger post-placement overload.
Ties favor lower $\mathrm{smax}_r(T_r/\bar T)+\mathrm{smax}_r(W_r/\bar W)$, then lower GPU index, with $\mathrm{smax}_r(x_r)=\lambda^{-1}\log\sum_r e^{\lambda x_r}$.
 
\paragraph{Local repair.}
Repair considers the highest-$\kappa_i$ sequences on the most and least loaded GPUs for each measure (Table~\ref{tab:buddyring_params}).
Local sequences can move from a high-load GPU to a low-load GPU or swap between the two.
A sequence on a high-load GPU can move to its parent group or, if split, to a same-size or child group containing a low-load GPU.
A group containing a marked GPU can move one sequence to each child.
Repair takes the best legal improvement until none remains or the iteration limit is reached.
 
\paragraph{Relaxation.}
If placement is infeasible, fillers enter the beam search by length class, longest first, trying group sizes 2, 4, then 8.
BuddyRing skips stages with unchanged groups. The final stage permits every legal group for every sequence.

\subsection{Critical-Wave Model Details}
\label{app:critical-wave-params}
 
\paragraph{Calibration.}
We calibrate the model in Section~\ref{sec:megadcp_critical_wave} using per-SM stage timings from Figure~\ref{fig:dcp-sm-utilization-motivation} (H100, decode-only, DCP degree 2).
Of 135 history-attention tasks, the first 132 map to CTAs in index order. The first three CTAs to finish each take one more task.
We grid-search the time unit, $\rho$, and $\delta_{\mathrm{attn}}$ to minimize root-mean-square error (RMSE) in per-CTA busy time.
For reduction, we schedule tasks on warps in queue order from observed CTA start times and fit $\delta_{\mathrm{red}}$ and $\gamma$ to sorted CTA busy times by least squares.
We use the fitted parameters (Table~\ref{tab:critical_wave_params}) across all evaluated DCP degrees and workloads.
 
\begin{table}[htb]
  \centering
  \vspace{-5pt}
  \caption{Critical-wave model parameters measured on H100. Overheads and costs are in model time units.}
  \label{tab:critical_wave_params}
  \begin{tabular}{@{}lr@{}}
    \toprule
    Parameter & Value \\
    \midrule
    Time per KV tile with all compute CTAs active & 1.08~$\mu$s \\
    Per-tile time floor $\rho$ & 0.43 \\
    History-attention overhead $\delta_{\mathrm{attn}}$ & 32 \\
    History-reduction overhead $\delta_{\mathrm{red}}$ & 2.5 \\
    Cost per merged partial vector $\gamma$ & 0.47 \\
    \bottomrule
  \end{tabular}
\end{table}
 
\paragraph{Fit quality.}
On the calibration batch, per-CTA RMSE is 1.2~$\mu$s for attention (measured range 42--67~$\mu$s) and 0.8~$\mu$s for reduction (16--35~$\mu$s).
The three CTAs with two attention tasks finish in 64.9--67.3~$\mu$s, compared with the predicted 65.5--66.4~$\mu$s.
The fixed cost of 32 tile-equivalents per attention task exceeds the 5--14 KV tiles each task processes in this batch.
 
\paragraph{Split planning.}
With $K$ compute CTAs, decode requests with at most 16 query tokens and at least $K$ local history KV tiles receive a split limit of at least four.
Score ties favor earlier attention completion, then fewer attention tasks, then fewer total splits.
The planner also uses $n_t+\delta_{\mathrm{attn}}$ for chunk-attention tasks.
For unsplit sequences, reduction tasks copy runs of output vectors. With $h_q$ local query heads, MegaDCP chooses the largest run in $\{1,h_q,2h_q,4h_q,\min(8h_q,32)\}$ that leaves at least $0.8\times12K$ tasks for the $12K$ warps.

\paragraph{Release-aware ordering.}
In split decode-only plans, each history task divides its modeled cost equally among the 16-token query blocks it needs.
Blocks transfer in decreasing order of their summed costs, in epochs of $\max(1,\lfloor M/P\rfloor)$ transfers for $M$ communication CTAs and DCP degree $P$.
History tasks run in order of their last required block's epoch, then by decreasing $c_{\mathrm{attn}}$.

\subsection{Execution Engine Details}
\label{app:engine-details}

\paragraph{MegaRing KV publication.}
After storing a KV tile for all heads, a communication CTA probes consecutive slots from the committed prefix using acquire--release compare-and-swap (CAS).
A failed CAS makes the occupied slot's KV data visible. After inserting the physical index at $p$, the CTA raises the prefix count to at least $p+1$ with a release \texttt{atomicMax}.
Producer warps acquire-load the count before claiming work. Once transfers finish, communication CTAs synchronize locally and reuse shared memory for compute.

\paragraph{MegaDCP CTA configuration.}
Each 384-thread compute CTA has three warp groups. Per-thread register quotas change from $24/240/240$ for attention to $168/168/168$ for reduction.
The leader claims 12 history-reduction descriptors at once, one per warp. For final reduction, eight warps execute one descriptor at a time per CTA.

\paragraph{MegaDCP block transfers.}
Each 128-row history-attention tile waits for its 16-token query blocks, which communication CTAs pull in host order.
After writing the send buffer, history-reduction warps increment the counter for the destination rank and 16-token output block.
For a remote block, the last contributing warp sets a ready flag on the query owner's GPU with system-scope release ordering.
The owner's communication CTA acquire-reads the flag, pulls the output and LSE, then signals local receive completion.

\section{Additional Evaluation Details}
\label{app:eval_details}

\subsection{Workloads and Baselines}

\begin{figure}[htb]
    \centering
    \includegraphics[width=0.9\columnwidth]{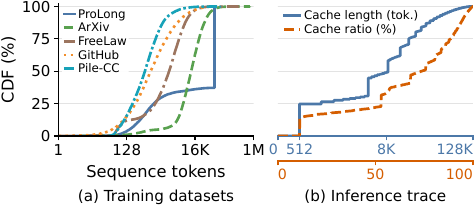}
    \vspace{-5pt}
    \caption{
        \textbf{Training and inference workload distributions.}
        (a) Sequence-length CDFs for five training datasets.
        (b) CDFs of cache length and cache ratio in the Mooncake trace.
    }
    \label{fig:workload-distributions}
\end{figure}

\begin{table}[htb]
  \centering
  \setlength{\tabcolsep}{1.pt}
  \vspace{-5pt}
  \caption{Mooncake snapshots by arrival-rate scale. D/M denotes decode-only/mixed batches. Request counts are means within each batch type. Lengths are means over request occurrences.}
  \label{tab:megadcp_trace_workload}
  \begin{tabular}{@{}lrrrr@{}}
    \toprule
    Arrival-rate scale & $1\times$ & $2\times$ & $4\times$ & All \\
    \midrule
    Snapshots (D/M) & 87/13 & 69/31 & 54/46 & 210/90 \\
    \midrule
    \multicolumn{5}{@{}l}{\emph{Mean requests per batch}} \\
    Decode-only: decode & 8.28 & 17.12 & 34.54 & 17.93 \\
    Mixed: decode & 8.38 & 17.26 & 34.26 & 24.67 \\
    Mixed: chunk prefill & 1.15 & 1.13 & 1.48 & 1.31 \\
    \midrule
    \multicolumn{5}{@{}l}{\emph{Mean length (tokens)}} \\
    Decode KV history & 12,156 & 12,479 & 12,174 & 12,259 \\
    Prefill KV history & 20,992 & 16,574 & 12,040 & 14,523 \\
    Prefill chunk & 2,420 & 3,486 & 2,772 & 2,939 \\
    \bottomrule
  \end{tabular}
\end{table}

We simulate Mooncake requests in 20\,ms steps using trace arrival times and input/output lengths. For each scale in Table~\ref{tab:megadcp_trace_workload}, we divide interarrival times by that factor and sample 100 snapshots between 300 and 3,600\,s.
Each decoding step accepts 0--4 of 15 drafts with probabilities $(0.20,0.25,0.25,0.20,0.10)$, committing 2.75 tokens on average including the target.
MegaDCP microbenchmarks use 32 query heads, head dimension 128, and TP degree 8. DCP degrees 2, 4, and 8 use four, two, and one KV heads, respectively, so each group shares one KV head.

The public mKernel~\cite{mkernel} (\texttt{68644c9}) and ParallelKittens/ThunderKittens~\cite{parallelkittens} (\texttt{34b15f7}) ring kernels lack the causal mask required for the MegaRing comparison.
vLLM uses more Triton-fused operators than SGLang. Against the best eager baseline, MegaDCP's geometric-mean speedups are 3.72$\times$ for decode-only and 1.52$\times$ for mixed batches, consistent with greater launch sensitivity for short queries.

\paragraph{Measurement protocol.}
MegaRing and MegaDCP run 10 and 40 warm-ups, respectively, before timing. CUDA Graph baselines add three eager calls before capture, then use graph replay. Uniform batches use the per-configuration communication-CTA sweep in Section~\ref{sec:eval_setup}.

\subsection{MegaRing Ablation Details}

The L1 baseline in Figure~\ref{fig:eval_megaring_ablation} uses eight attention launches, seven NCCL transfers, and eight external reductions.

Our Megatron baseline uses the hybrid-CP scheduler from Megatron-LM~\cite{megatron_hybrid_cp} (\texttt{368fa88}) to select power-of-two CP degrees from token targets, balance attention work, and expand groups onto idle ranks.
Our padded implementation of Zeppelin's Algorithm~2~\cite{zeppelin} allocates GPUs to long sequences in proportion to squared length and packs short sequences locally. Its linear modules retain the attention layout.

With the MegaRing kernel fixed, BuddyRing gains up to 1.16/1.18$\times$ in forward/backward throughput over Megatron-derived placement on FreeLaw. With placement fixed, MegaRing outperforms FA3/NCCL by 1.11--1.34$\times$ for Megatron and 1.62--2.13$\times$ for Zeppelin (Figure~\ref{fig:eval_megaring_load_balance}).

\subsection{MegaDCP Stage Profiling}
\label{app:eval_megadcp_timestamps}

\begin{figure}[htb]
  \centering
  \includegraphics[width=0.95\columnwidth]{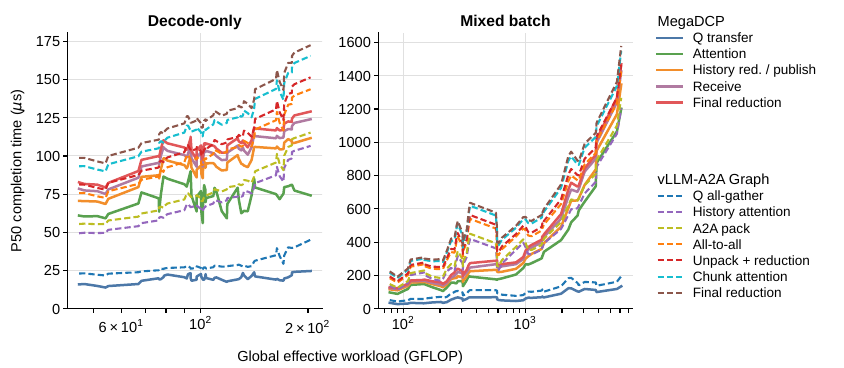}
  \par\vspace{1pt}
  {\small\setlength{\tabcolsep}{2pt}
  \begin{tabular*}{0.95\columnwidth}{@{\extracolsep{\fill}}lrrrr@{}}
    \toprule
    Batch & \shortstack{Query\\transfer} & Attention & \shortstack{Result\\exchange} & \shortstack{Final\\reduction} \\
    \midrule
    Decode-only & 20/29 & 73/71 & 102/109 & 107/128 \\
    Mixed & 67/104 & 297/376 & 376/506 & 394/570 \\
    \bottomrule
  \end{tabular*}}
  \vspace{-5pt}
  \caption{Stage completion at DCP degree 4 and $4\times$ arrivals. Entries give mean times ($\mu$s) for MegaDCP/vLLM-A2A Graph. Exchange includes history reduction. Both use a common host-side start. MegaDCP (solid) timestamps the last task's completion in each stage in-kernel and adds launch latency. vLLM-A2A Graph (dashed) accumulates CUDA-event durations scaled per case to latency without profiling events.}
  \label{fig:eval_megadcp_timestamp}
\end{figure}

Figure~\ref{fig:eval_megadcp_timestamp} orders cases by total effective attention FLOPs across GPUs and smooths per-case medians with a $1:2:1$ moving average. These cross-system timings reflect overlap alongside other implementation differences. MegaDCP is faster in 53 of 54 decode-only cases and all 46 mixed cases. Mean mixed-batch speedups of 1.81$\times$ in the lowest FLOP quartile and 1.20$\times$ in the highest are consistent with less scope for overlap as attention dominates.

\subsection{Uniform-Length MegaRing Results}
\label{app:eval_uniform_megaring}

BuddyRing and Zeppelin assign each $T/B$-token sequence to the same $8/B$ GPUs (Table~\ref{tab:megaring_uniform}). Their comparison isolates execution differences, including fused reductions and readiness-driven coalescing.

\begin{table}[htb]
\centering
\vspace{-5pt}
\caption{MegaRing Hybrid forward/backward speedups ($\times$) for $B$ equal-length sequences totaling $T$ tokens. Baselines, in column order, are all-gather, Llama-3 style, Ring FA, Megatron, MagiAttention, and Zeppelin.}
\label{tab:megaring_uniform}
\begingroup
\fontsize{8}{10}\selectfont
\setlength{\tabcolsep}{1.5pt}
\begin{tabular*}{\columnwidth}{@{\extracolsep{\fill}}rr*{6}{c}@{}}
\toprule
$T$ & $B$ & AG & L3 & Ring & Meg. & Magi & Zep. \\
\midrule
64K & 1
& 1.05/1.22 & 1.03/1.15 & 1.26/1.27
& 1.29/1.28 & 1.06/1.31 & 1.29/1.28 \\
64K & 2
& 1.20/1.49 & 1.14/1.29 & 1.85/1.84
& 1.13/1.11 & 1.16/1.29 & 1.13/1.11 \\
64K & 4
& 1.53/2.03 & 1.38/1.51 & 3.06/2.47
& 1.07/1.03 & 1.34/1.31 & 1.09/1.03 \\
\midrule
128K & 1
& 1.00/1.10 & 1.02/1.06 & 1.09/1.13
& 1.11/1.14 & 1.00/1.19 & 1.12/1.13 \\
128K & 2
& 1.04/1.26 & 1.06/1.16 & 1.30/1.38
& 1.31/1.32 & 1.03/1.17 & 1.04/1.05 \\
128K & 4
& 1.23/1.50 & 1.17/1.24 & 1.80/1.84
& 1.18/1.15 & 1.12/1.14 & 1.10/0.99 \\
\midrule
256K & 1
& 1.02/1.02 & 1.00/1.02 & 1.02/0.99
& 1.02/1.00 & 1.01/1.10 & 1.02/1.00 \\
256K & 2
& 1.02/1.11 & 1.02/1.08 & 1.11/1.10
& 1.13/1.15 & 0.99/1.10 & 0.99/1.00 \\
256K & 4
& 1.10/1.25 & 1.10/1.13 & 1.36/1.26
& 1.37/1.34 & 1.21/1.10 & 1.00/0.97 \\
\bottomrule
\end{tabular*}
\endgroup
\end{table}

Against the fastest baseline per configuration, Hybrid's geometric-mean speedup is 1.04$\times$ in both directions (forward 0.99--1.13$\times$, backward 0.97--1.15$\times$).
At 64K (8K tokens per GPU), per-step transfers and merges cost proportionally more. For $B=4$, Ring FA's eight-way split of each 16K sequence also makes placement a factor. For a single 128K or 256K sequence, all methods use eight GPUs and the fastest baselines, mostly all-gather variants, hide communication behind attention (Hybrid speedup 0.99--1.06$\times$).

\subsection{Uniform-History MegaDCP Results}
\label{app:eval_uniform_megadcp}

Against vLLM-A2A Graph, the strongest trace baseline, MegaDCP achieves 1.24$\times$ on 24 decode-only configurations and 1.20$\times$ on 48 mixed configurations (geometric means). The respective ranges are 1.03--1.37$\times$ and 1.05--1.53$\times$ (Table~\ref{tab:megadcp_uniform}).

\begin{table}[htb]
\centering
\vspace{-5pt}
\caption{MegaDCP speedups ($\times$) over vLLM-A2A Graph at DCP degrees 2/4/8 for history length $L_{\mathrm{KV}}$ per request. Each batch has $B_d$ decode requests with 16 query tokens each and $B_p$ prefill requests with one 2K-token chunk each.}
\label{tab:megadcp_uniform}
\begingroup
\fontsize{8}{10}\selectfont
\setlength{\tabcolsep}{1.pt}
\begin{tabular*}{\columnwidth}{@{\extracolsep{\fill}}rr*{4}{c}@{}}
\toprule
\multicolumn{2}{c}{Workload}
& \multicolumn{4}{c}{$L_{\mathrm{KV}}$} \\
\cmidrule(lr){1-2}
\cmidrule(lr){3-6}
$B_p$ & $B_d$ & 16K & 32K & 64K & 128K \\
\midrule
0 & 4
& 1.25/1.35/1.35 & 1.22/1.32/1.36
& 1.26/1.37/1.32 & 1.24/1.28/1.30 \\
  & 32
& 1.34/1.29/1.37 & 1.24/1.16/1.15
& 1.15/1.14/1.03 & 1.10/1.11/1.09 \\
\midrule
1 & 4
& 1.05/1.40/1.41 & 1.08/1.14/1.30
& 1.07/1.15/1.22 & 1.17/1.18/1.07 \\
  & 32
& 1.42/1.53/1.38 & 1.27/1.38/1.29
& 1.15/1.24/1.23 & 1.12/1.11/1.07 \\
\midrule
2 & 4
& 1.15/1.33/1.35 & 1.17/1.33/1.24
& 1.07/1.12/1.13 & 1.08/1.06/1.06 \\
  & 32
& 1.42/1.36/1.35 & 1.27/1.25/1.23
& 1.16/1.16/1.12 & 1.10/1.08/1.07 \\
\bottomrule
\end{tabular*}
\endgroup
\end{table}

In decode-only batches of 32 requests, speedups fall from 1.29--1.37$\times$ at 16K to 1.09--1.11$\times$ at 128K as attention dominates. With four requests, history splitting exposes parallelism across all lengths (1.22--1.37$\times$).
Across all 72 configurations, geometric-mean speedups over vLLM-AG-RS Graph and SGLang-AG-AR Graph are 1.30$\times$ and 1.70$\times$, respectively.

\subsection{End-to-End Evaluation Details}
\label{app:eval_end_to_end}

Both experiments use Qwen3-30B-A3B~\cite{qwen3} (32 query heads, four KV heads, head dimension 128).

\paragraph{Transformer-layer training.}
Attention uses eight-way CP and the MoE block uses eight-way expert parallelism (EP). All weights, including the expert router, are randomly initialized.
In Figure~\ref{fig:eval_megaring_end2end}, \emph{Core Attention} includes CP communication. \emph{Other} covers projections, EP communication, and remaining operations. On ProLong, \emph{Other} latency for Megatron, Zeppelin, and MagiAttention is close to uniform partitioning. It rises with token imbalance on short-sequence-heavy workloads. Both MegaRing variants stay close to uniform partitioning.

\paragraph{vLLM serving.}
We replay a contiguous 303\,s Mooncake segment~\cite{mooncake} (100 warm-up and 1,000 measured requests). Only new prompt tokens need chunked prefill on the decode side of disaggregated serving. With TP and EP degrees of 8 and DCP degree 2, each rank pair shares one KV head.
Figure~\ref{fig:eval_megadcp_end2end} varies arrival-rate scale (1, 2, 4) and maximum batch size (64, 128). Speedups relative to the best vLLM baseline per configuration and metric are averaged arithmetically.

At $4\times$ arrivals and batch size 128, attention takes a larger share of each iteration and TBT speedup reaches 1.10$\times$. At batch size 64, faster iterations shorten the wait for new prompts. With the shared engine, MegaDCP naive reaches up to 1.07$\times$ for TBT and 1.29$\times$ for TTFT before critical-wave planning and release-aware ordering.

\end{document}